\documentclass{aa}  

\usepackage{graphicx}
\usepackage{txfonts}
\usepackage{subcaption}         
\usepackage{lscape}             
\usepackage{placeins}           
\usepackage{natbib}
\usepackage{amssymb,amsmath}
\bibpunct{(}{)}{;}{a}{}{,}
\usepackage[hidelinks]{hyperref}
\usepackage{caption}
\usepackage{float}
\usepackage{xcolor}
\renewcommand{\thesubfigure}{(\alph{subfigure})}
                                
\begin{document}

\title{Reducing False Positives in Strong-Lens Searches with Generalized-Mean Consensus of Machine-Learning Ensembles in the Kilo-Degree Survey}
\titlerunning{Reducing False Positives in Strong-Lens Searches}

\author{Ziqi Li\inst{1,2},
        Rui Li\inst{1} \fnmsep\thanks{Corresponding author: liruiww@gmail.com},
        Xu Huang\inst{1},
        Hui Li\inst{1},
        Pufan Liu\inst{1,2},
        Liang Gao\inst{1,3},
        Crescenzo Tortora\inst{4},
        Nicola N. Napolitano\inst{5},
        Xiaoyue Cao\inst{1},
        Ran Li\inst{3},
        Liqing Chen\inst{1},
        Kang Jiao\inst{1},
        Valerio Busillo\inst{4},
        Yue Dong\inst{6}}
        
 \institute{$^{1}$Institute for Astrophysics, School of Physics, Zhengzhou University, Zhengzhou, 450001, China.\\
   $^{2}$International College, Zhengzhou University, Zhengzhou, 450001, China \\
   $^{3}$School of Physics and Astronomy, Beijing Normal University,  Beijing 100875, China.\\
   $^{4}$INAF – Osservatorio Astronomico di Capodimonte, Salita Moiariello 16, I-80131, Napoli, Italy.\\
   $^{5}$Department of Physics ``E. Pancini'', University Federico II, Via Cinthia 6, 80126-I, Naples, Italy.\\
   $^{6}$School of Mathematics and Physics, Xi'an Jiaotong-Liverpool University, 111 Renai Road, Suzhou, 215123, People's Republic of China.}

\date{}
 
  \abstract
   {In wide-field surveys, the main challenge for identifying galaxy-galaxy strong lenses is not just classifier sensitivity, but the overwhelming number of false positives. Searching for rare systems like strong lenses among millions to bilions of galaxies inevitably produces many contaminants, making this the primary bottleneck for follow-up inspection and for building statistically useful lens samples.}
   { We aim to improve the purity of strong-lens candidate selection in KiDS DR4 by combining several independently classifiers rather than relying on a single network architecture. The objective is to retain high completeness for known candidates while substantially reducing the fraction of non-lenses.}
   {We trained a set of convolutional, Transformer-based, and hybrid classifiers, including Li ResNet+, Swin Transformer variants, Swin-MLP, and DemiLensNet. Their probabilistic outputs were combined at the score level using simple averaging and a generalized mean consensus. The models were first tested on simulated KiDS-like lens images and then evaluated on real KiDS DR4 lens candidates embedded in a non-lens sample.}
   {On the simulated test set, ensembles show no clear advantage over the best single models.
    On the mixed real KiDS test set, however, the arithmetic mean reduces the false-positive rate at 90\% completeness from 0.016--0.020 (the range spanned by the two best individual models) to 0.011 for the seven-model ensemble.
    The generalized mean reduces it further, to 0.007.
    Applied to the full LRG and BG samples of KiDS DR4 at the same 90\% completeness level, the generalized mean strategy reduces the number of returned candidates by roughly 50\% for LRGs and 70\% for BGs, relative to the best single model.
    This reduction brings the candidate list within reach of systematic visual inspection.
    After visual inspection and classification, we obtain 170 new high-quality candidates (24 Class~A and 146 Class~B), together with 1706 Class~C candidates.}
   {Our results demonstrate that the generalized mean consensus of an ML ensemble strategy provides a practical route to reducing the visual inspection workload while preserving a high recovery rate of promising strong-lens candidates.}

   \keywords{strong gravitational lensing, deep learning, convolutional neural network, Transformer, hybrid architecture, image classification}
   \authorrunning{Z. Li et al.}
   \maketitle
   \nolinenumbers

\section{Introduction}
Strong gravitational lensing is an essential tool for studying the mass distribution in galaxies and clusters, and can reveal the structure and evolution of distant galaxies (e.g., \citealt{2009astro2010S.159K, Sonnenfeld2015ApJ...800...94S, Shajib2021MNRAS.503.2380S, vanDokkum2024NatAs...8..119V, 2023MNRAS.521.6005E, Li2018MNRAS.480..431L, Li2025ApJ...987L..31L}) as well as key cosmological parameters (e.g., \citealt{Suyu2013ApJ...766...70S, Wong2020MNRAS.498.1420W, Chen2019MNRAS.488.3745C, Li2023PDU....4101234L, 2024SSRv..220...48B}). Realizing this potential requires building large, well-understood lens samples. The key bottleneck is not only confirming and modeling lenses, but also discovering promising candidates within the vast number of galaxy images. As survey volumes grow, neither human inspection nor expensive follow-up observations can keep up. Therefore, lens-finding algorithms must operate efficiently with limited computational and human resources while maintaining high scientific yield and reliability. During candidate searching, a search tuned only to maximize its overall score can still produce a long list of false positives, which directly increases the review workload.

In third-generation surveys, the typical workflow consists of a probabilistic classifier, a selection threshold to define a manageable candidate set, and human inspection and/or additional checks for confirmation (\citealt{Petrillo2019MNRAS.484.3879P, 2019ApJS..243...17J, Huang2021ApJ...909...27H, li2020, Wong2022PASJ...74.1209W, Shu2022A&A...662A...4S, He2025A&A...695A..76H}). At a fixed completeness, the number of surviving candidates is governed by the remaining false-positive fraction and the total number of objects being classified. When the screened sample reaches millions, even a small residual false-positive fraction at a given completeness threshold (e.g., 90\%) can yield tens of thousands of candidates (e.g. \citealt{li2021, 2019ApJS..243...17J}), making the bottleneck increasingly about keeping the shortlist small enough for reliable verification. Under this regime, reducing the tail of high-score non-lenses can produce disproportionately large practical gains.
Fourth-generation surveys will exacerbate this issue. The number of screened galaxies is expected to increase by roughly one to two orders of magnitude, with broader and deeper imaging (e.g., \citealt{Collett_2015, Cao2024MNRAS.533.1960C,2026A&A...711A..27E,2026A&A...711A..28E}). If the completeness requirement is unchanged, the absolute number of false positives scales with the residual false-positive fraction at the selected threshold, quickly outpacing verification capacity. In the future, although this task is often described as "lens detection," for large surveys it is more accurately framed as catalogue filtering. The objective is to produce a shortlist of strong-lens candidates for efficient follow-up while retaining known lenses at a specified completeness.

Prior work has largely focused on improving a single model's ability to separate lenses from non-lenses, including better network designs and training strategies for robustness to image quality, seeing, and blending (\citealt{Petrillo2019_test}). However, systematic confusion can never be completely avoided. Spiral arms, interacting systems, or compact blue sources may share image features with lens arcs after convolution with the point-spread function, and residual artifacts can produce localized curvature-like structures. When such patterns induce falsely high scores through the features learned by the model, retuning thresholds alone may not eliminate the dominant contribution from the tail of false positives. This motivates a second route: combining multiple models with different built-in preferences and merging their outputs. Model combinations can improve performance on new data and are often implemented as averaging or voting of predicted probabilities (e.g., \citealt{Rezaei2025}).
However, simple averaging or voting is not sufficient. The critical issue is how the combined model set handles rare but high-impact false positives that appear near the top of the ranked list at a fixed completeness level. Simple averaging and voting can still be dominated by a small subset of models that assign very high probabilities to the same non-lens morphology. These observations motivate a targeted use of model diversity: constructing a combination rule that suppresses the influence of outliers in the high-score tail while allowing high final scores only when models agree robustly. We therefore introduce a generalized-mean combination (defined below) that places greater emphasis on agreement across models. Compared with the standard mean or vote, the generalized mean reduces the weight of unusually high predictions that do not receive consistent support across the model set, with the goal of shortening the score tail that drives false positives at a given completeness threshold, thereby improving the candidate shortlist without sacrificing completeness.

We build a set of independently trained models with different built-in pattern preferences and image-processing schemes: convolutional networks emphasize compact arc-like structures, Transformer-based models capture broader context, and hybrid networks combine local and global representations. While each model is trained to distinguish lenses from non-lenses, they do not make the same kinds of mistakes. In addition to the arithmetic mean, we use a generalized-mean combination, choosing a merging scheme in which low probability values from individual models pull the final score downward, while a single high outlier cannot lift the score on its own. As a result, a candidate receives a high final score only when multiple models assign consistently high probabilities, rather than when one classifier produces an extreme value for a particular non-lens morphology. The arithmetic mean provides a direct average of the model outputs, whereas the generalized mean imposes a stronger requirement for cross-model agreement. The combined model output is therefore not used to maximize overall accuracy, but to reduce false-positive contamination of the candidate set while preserving the completeness required for a practical survey.

The remainder of this article is organized as follows. In Sect.~\ref{sec:data}, we describe the dataset preparation, including simulation of lens images and the composition of training and test sets. Sect.~\ref{sec:methods} details the DemiLensNet architecture and the baseline models, as well as the training procedure and evaluation metrics. Sect.~\ref{sec:results} presents the experimental results. In Sect.~\ref{sec:discussion}, we discuss the implications of these results for future surveys. Finally, Sect.~\ref{sec:conclusion} summarizes our findings and concludes the paper.

\section{Data}
\label{sec:data}

All data preparation and all lens simulations in this study follow \cite{li2021} without changes to the method or parameter choices. We adopt their full pipeline so that our results are directly comparable. In short, we create multi-band image cutouts, add simulated lensed arcs on top of real foreground galaxies, apply a band-dependent point-spread function (PSF), and enforce the same visibility cuts. We also assemble a large and diverse set of negative examples. The goal is to reproduce these data properties and selection criteria as closely as possible, while using the sample sizes needed for this work.

We use image cutouts in the $g$, $r$, and $i$ bands. Each cutout covers \(20^{\prime\prime}\times20^{\prime\prime}\) on a \(101\times101\) pixel grid, matching the adopted sampling and field of view. The background and noise levels follow the same treatment. We keep the true light of the foreground galaxy in each cutout and only add simulated features for the lensed source. This preserves the shape and color of the lens galaxy in a realistic way. The foreground sample consists of \(30{,}000\) luminous red galaxies (LRGs) that act as lens hosts. Every one of these LRGs was inspected by eye. We confirm that none shows lens-like arcs or rings, and none suffers from clear artifacts such as star trails, satellite tracks, or bright halos. 

We add the simulated lensed arcs to these real LRG images after PSF convolution in each band, and we match the noise and background to the data. These steps ensure a realistic blend of lens and source light. The background (lensed) sources follow the design and parameter distributions of \cite{li2021}. The surface brightness profile is Sérsic. Source redshifts and apparent magnitudes are drawn from a deep catalog consistent with Rubin/LSST; we use \(0.8<z<3.0\) and \(21<r<25\). The $g$, $r$, and $i$ colors come from the same catalog, and we add a small random jitter of \(\pm 0.1\) mag per band to reflect intrinsic color scatter. Structural parameters follow the same distributions: the effective radius \(R_{\mathrm{eff}}\) is drawn from a normal distribution with mean \(0.2^{\prime\prime}\) and standard deviation \(0.3^{\prime\prime}\), truncated to \(0.1^{\prime\prime}\)–\(0.5^{\prime\prime}\); the Sérsic index \(n\) is uniform in the range 0.3–5; the axis ratio and position angle are drawn from uniform distributions.

The lens mass model is a singular isothermal ellipsoid (SIE). The Einstein radius \(R_{\mathrm{Ein}}\) is drawn from an exponential distribution over 1–5 arcsec. We include external shear with amplitude \(\gamma\) uniform in 0–0.1 and shear angle uniform in \(0^{\circ}\)–\(180^{\circ}\). To capture small-scale complexity, we add a Gaussian random field (GRF) perturbation to the lens potential with a fixed power-law slope of \(-6\). The potential variance is drawn from a log-uniform range \(10^{-4}\)–\(10^{-1}\). These terms introduce deviations from a smooth lens and produce arcs with more realistic shape and substructure. Multi-band ray tracing then yields the lensed source in $g$, $r$, and $i$, keeping the morphology and color of the source consistent across bands. We model the PSF in each band with a Moffat profile with \(\beta=2.2\), and draw the full width at half maximum (FWHM) to match the adopted KiDS seeing statistics. In the g band, the FWHM is normal with mean \(0.85^{\prime\prime}\) and standard deviation \(0.10^{\prime\prime}\), clipped to 0.60–1.20\(^{\prime\prime}\). In the r band, the FWHM is normal with mean \(0.70^{\prime\prime}\) and standard deviation \(0.05^{\prime\prime}\), clipped to 0.50–0.90\(^{\prime\prime}\). In the i band, the FWHM is normal with mean \(0.80^{\prime\prime}\) and standard deviation \(0.10^{\prime\prime}\), clipped to 0.55–1.20\(^{\prime\prime}\). We convolve the simulated source light with the band PSF before adding it to the real foreground image. We then adjust noise and background to the data cutout. As in Li et al.\ (2021), we impose visibility cuts to control purity and to remove cases with very weak or confused arcs. We keep a system if the peak pixel value of the arc divided by the r-band peak of the lens is at least \(\alpha/\beta \ge 0.05\), or if the local peak within a \(3\times3\) pixel window has signal-to-noise \(>5\). These filters remove very low-contrast or very small-separation arcs. This reduces completeness at the faint end but raises the purity of the final set.

The negative sample follows the same philosophy. We assemble \(45{,}000\) negatives. Of these, \(30{,}000\) are the LRGs act as the lens hosts above. The rest include a wide range of galaxy types that can mimic arc-like shapes, such as spiral galaxies with clear arms or rings, interacting and merging systems, edge-on disks, irregular galaxies, and systems with bright star-forming knots. We check that none of these negatives shows a lens-like arc and that strong artifacts or saturation are absent. This produces a challenging but clean control set.

We label all simulated lenses as class 1 (positive) and all contaminants as class 0 (negative). The dataset comprises a total of 80,000 individual images (45,000 positives and 45,000) and is split following standard machine learning paradigms into training, validation, and test sets. The training set contains 74,000 images (randomly shuffled, accounting for $\sim$ 82\% of the total) and is used for model parameter optimization. The validation set consists of 8,000 images 9\%) and is used to monitor the training process and guide hyperparameter tuning. The test set includes 8,000 images (9\%) and is used for final evaluation of the model's generalization performance.

\section{The Classification Models}
\label{sec:methods}

\subsection{Models with Diverse Architectures}
\label{sec:model}

\begin{figure*}[h!]
\centering
\includegraphics[width=0.8\textwidth]{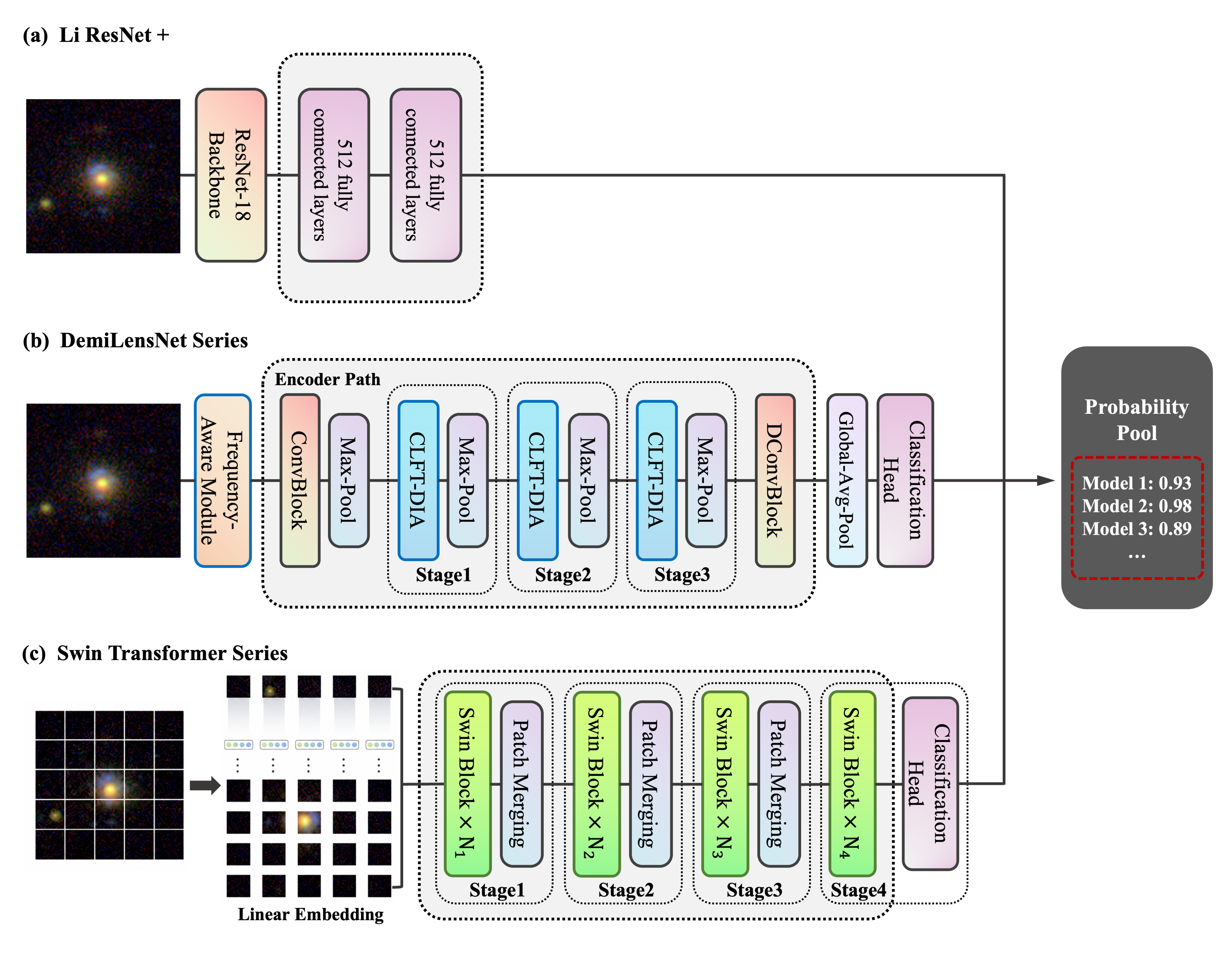}
\caption{Schematic of the model architectures.
(a) Li ResNet+.
(b) DemiLensNet series.
(c) Swin Transformer series, where $N_1$--$N_4$ indicate the number of blocks in the four Swin stages.}
\label{fig:model}
\end{figure*}

We construct a heterogeneous set of image-classification models covering three complementary design principles: (1) convolutional networks with residual connections, (2) hybrid architectures that fuse convolutional feature extraction with self-attention modules, and (3) hierarchical vision Transformers alongside their MLP-based counterparts. Figure~\ref{fig:model} presents an overview of the three model groups used in this work.

\textbf{Li ResNet+.}
Li ResNet+ is a ResNet-based image classifier used in Li et al.~\cite{li2021}. It is based on ResNet-18, in which residual shortcut connections help train deeper convolutional networks more reliably \cite{he2016deep}. Compared with the standard ResNet-18 classifier, Li ResNet+ adds two fully connected layers, each with 512 neurons, before the final lens/non-lens classification layer. This gives the classifier more flexibility in the final classification step while leaving the main ResNet-18 convolutional structure unchanged. We train this model using the same preprocessing, loss function, and validation protocol as for the other models.

\textbf{DemiLensNet Series.}
We use two variants of DemiLensNet: DemiLensNet and DemiLensNet-L. 
Both variants share the same overall architecture, while DemiLensNet-L is the larger version, increasing the backbone feature dimension from \texttt{dim=32} to \texttt{dim=64} and containing approximately 40.21 million parameters. 
The architecture is based on an encoder--decoder framework adapted from CLFTNet \cite{abc2023}. 
In this framework, convolutional branches extract local structures such as edges and textures, while CLFT modules combine these local features with broader contextual information. 
This combination is useful for detecting faint arc-like features, which are often spatially extended but have low surface brightness locally.
At shallow stages, DemiLensNet includes a Frequency-Aware (FA) module to reduce small-scale noise before repeated downsampling. The FA module combines frequency-channel attention, following FcaNet \cite{fca2021}, with a learnable Fourier filter. These operations help suppress noisy high-resolution features while preserving structures that resemble lensing arcs.
At deeper stages, DemiLensNet inserts Deformable Interactive Attention (DIA) modules \cite{dia2025} into the CLFT blocks, forming CLFT-DIA units. These modules combine broad contextual information with localized high-contrast features and allow the attention map to better follow extended or irregular structures. Within each CLFT-DIA unit, the CLFT output and the attention-enhanced representation are combined through learned weights and stabilized by a residual connection. 
The final representation is passed through global average pooling and a lens/non-lens classification layer. 
For the base DemiLensNet with \texttt{dim=32}, we add a 512--256 fully connected transformation before the output layer to increase the flexibility of the final classifier. 
For DemiLensNet-L, this extra transformation is omitted because the larger backbone already provides a higher-dimensional representation.

\textbf{Swin Transformer Series.}
The Swin Transformer series includes Swin-T, Swin-S, Swin-B, and Swin-MLP.
These models follow the Vision Transformer (ViT) approach \cite{dosovitskiy2021imageworth16x16words}, but replace global self-attention over all image patches with a hierarchical shifted-window design \cite{2021arXiv210314030L}.
An input image is first divided into non-overlapping patches and embedded as visual tokens.
The backbone then processes these tokens through multiple stages, where patch-merging layers progressively reduce spatial resolution and increase the number of channels.
Within each stage, window-based self-attention captures relationships among nearby patches, while shifted windows link patches across adjacent windows in successive blocks.
This design is less computationally expensive than global attention and provides a multi-scale representation suitable for compact astronomical images.
Swin-T, Swin-S, and Swin-B all use the same shifted-window principle but differ in width, depth, and overall capacity.
Swin-T is the lightest configuration, Swin-S offers intermediate capacity, and Swin-B is the largest variant.
Swin-MLP retains the same hierarchical shifted-window structure but replaces the self-attention operation with a simpler MLP-based mixing module.
This gives Swin-MLP a different inductive bias from the attention-based variants and contributes to the diversity of the ensemble.
For all Swin variants, the original ImageNet classification head is replaced by a lens/non-lens classification layer, and the models are trained with the same data split and evaluation pipeline as the Li ResNet+ and DemiLensNet models.

\subsection{Training the models}
This work was conducted using Python and the PyTorch deep learning framework. All models are trained under the NVIDIA A100 GPU environment. We use the AdamW optimizer (Adam with decoupled weight decay) for optimization (\citealt{liu2021}), with an initial learning rate of $5 \times 10^{-4}$ and a weight decay of $10^{-5}$. The binary classification loss function is Binary Cross-Entropy with Logits (BCEWithLogitsLoss) (\citealt{rojas2022}). A cosine annealing learning rate scheduler (CosineAnnealingLR) is applied (\citealt{liu2021}), with a cycle period of $T_{\text{max}} = 30$ and a minimum learning rate $\eta_{\min} = 10^{-11}$. The maximum number of training epochs is set to 500, and we adopt a “save best” strategy to avoid overfitting. For pure Transformer-based architectures (Swin Transformer and ViT), we apply a 10-epoch linear warmup to accelerate convergence (\citealt{dosovitskiy2021}). For DemiLensNet and its variants, we use fewer warmup steps.

The input consists of three-channel ($g$, $r$ and $i$ bands) PNG images, center-cropped to 96×96 pixels (this scale has been verified to encompass all simulated lens structures). During the loading stage, each pixel value is scaled from the 0–255 \texttt{uint8} range to a [0, 1] \texttt{float32} range for normalization. In the early phase of training, we apply data augmentation techniques including random resizing (from 96 to 105.6 pixels), random cropping, horizontal/vertical flipping, 180° rotation, and 5\% probability grayscale conversion. 
After the validation performance plateaus, we disable all data augmentation and continue training on the original images. This final unaugmented stage is motivated by evidence that data augmentation has limited additional regularization effects near convergence \citep{golatkar2019time}.

Regarding noise handling, since the simulated data already contains realistic noise, no additional noise is introduced during training. As color carries astrophysically relevant information and strong-lens classifiers can be highly sensitive to it, we do not apply color jittering \citep{2022A&C....3800535J}. Dataset construction is subject to strict quality control, combining manual inspection with automated scripts to ensure the physical validity of lens samples and the accuracy of their labels.

\section{Performances}
\label{sec:results}
In this section, we assess the behavior of the models as individual classifiers and when their outputs are combined into an ensemble.
Since the intended application is not balanced-set classification but candidate filtering in wide-field surveys, we focus on the regime where false positives must be kept low.
This is especially important for strong-lens searches: because true lenses are rare, even a modest contamination rate can dominate the workload of visual inspection and follow-up observations \citep{Collett_2015, li2020, li2021, denselens2023, 2025MNRAS.538.1081R}.

\subsection{Evaluation metrics}

We evaluate classification performance for each model with four standard metrics: accuracy, area under the receiver operating characteristic curve (AUC), true positive rate (TPR), and false positive rate (FPR).
Accuracy is the fraction of all objects that are correctly classified:
\begin{equation}
\mathrm{Accuracy} = \frac{\mathrm{TP} + \mathrm{TN}}{\mathrm{TP} + \mathrm{TN} + \mathrm{FP} + \mathrm{FN}},
\end{equation}
where TP, FN, FP, and TN denote the numbers of true positives, false negatives, false positives, and true negatives, respectively.
The AUC summarizes the trade-off between true and false positive rates across all possible decision thresholds and provides a single global measure of discriminative power.
The TPR, also known as completeness, is the fraction of true lenses that are correctly identified:
\begin{equation}
\mathrm{TPR} = \frac{\mathrm{TP}}{\mathrm{TP} + \mathrm{FN}},
\end{equation}
and the FPR is the fraction of non-lenses that are misclassified as lenses:
\begin{equation}
\mathrm{FPR} = \frac{\mathrm{FP}}{\mathrm{FP} + \mathrm{TN}}.
\end{equation}
While accuracy and AUC are useful summary statistics, they do not fully capture what matters in the lens-search setting.
Because true lenses are extremely rare among the millions of galaxies in a wide-field survey, even a small FPR can produce far more false positives than true candidates.
Our primary goal is therefore to achieve an FPR as low as possible while maintaining high TPR.
Which metric is most informative depends on the composition of the test sample.
For the balanced simulated test set, we report TPR at a fixed $\mathrm{FPR}=0.5\%$. This metric probes model behavior under strict contamination control and is more informative than accuracy alone when AUC values are all close to unity.
For the real-data benchmark, where the sample is dominated by non-lenses, we instead report $\mathrm{FPR}$ at $\mathrm{TPR}=90\%$, which directly measures how many false positives remain when most true lenses are recovered.
Both AUC and the relevant threshold-based metric are reported for each model.
The decision thresholds listed in Table~\ref{tab:comparison} correspond to these summary statistics.

With the individual model performance established, we now describe how their outputs are combined into ensembles.
We construct ensembles by progressively adding models in order (from best to weakest) of their individual performance on the simulated test set (see Sect.~\ref{sec:performance_comparison}). We evaluate two ways of combining the predicted probabilities.
The first is the simple arithmetic mean,
\[
P_{\mathrm{ens}} = \frac{1}{N}\sum_{i=1}^{N} p_i .
\]
The second is the generalized mean,
\[
P_{\mathrm{ens}} =
\left(
\frac{1}{N}\sum_{i=1}^{N}p_i^{\,r}
\right)^{1/r},
\]
where $p_i$ is the predicted lens probability from the $i$-th model and $r$ is the order of the generalized mean. Arithmetic averaging ($r=1$) has been explored in strong-lens ensembles \citep{2023MNRAS.523.4188N}, while \citet{2024MNRAS.530.1297H} additionally tested harmonic averaging ($r=-1$), a particular case of generalized mean. In this work we adopt $r=-0.5$, chosen to minimise the number of LRG candidates at 90\% completeness on the real-data benchmark (see details in Sect.~\ref{sec:new_candidates}, also applied in test data). 
This value makes the ensemble sensitive to low-confidence predictions: if any model assigns a low probability, the generalized mean is pulled downward.
This acts as a stricter consensus filter that only promotes candidates for which most models agree on a high score.

\subsection{Performance on Simulated Data}
\label{sec:performance_comparison}

\begin{table*}
\caption{Performance comparison for the seven models on simulated data.}
\label{tab:comparison}
\centering
\renewcommand{\arraystretch}{1.25}%
\begin{tabular}{lcccccc}
\hline\hline
Model & Accuracy & TPR & FPR & Threshold & TPR@FPR=0.5\% & AUC\\
\hline
\multicolumn{7}{c}{Single model}\\
\hline
DemiLensNet             & 0.9870 & 0.9850 & 0.0103 & 0.4412 & 0.9743 & 0.9989\\
Li ResNet+              & 0.9817 & 0.9837 & 0.0180 & 0.3333 & 0.9593 & 0.9982\\
DemiLensNet-L           & 0.9860 & 0.9783 & 0.0053 & 0.8109 & 0.9770 & 0.9984\\
Swin-MLP                & 0.9805 & 0.9800 & 0.0180 & 0.3806 & 0.9530 & 0.9978\\
Swin-Tiny               & 0.9792 & 0.9787 & 0.0180 & 0.2677 & 0.9580 & 0.9978\\
Swin-Small              & 0.9785 & 0.9743 & 0.0157 & 0.3998 & 0.9547 & 0.9969\\
Swin-Base               & 0.9798 & 0.9733 & 0.0117 & 0.5852 & 0.9533 & 0.9975\\
\hline
\multicolumn{7}{c}{Arithmetic mean}\\
\hline
2 models & 0.9858 & 0.9807 & 0.0080 & 0.5529 & 0.9717 & 0.9988\\
3 models & 0.9875 & 0.9823 & 0.0070 & 0.5156 & 0.9757 & 0.9989\\
4 models & 0.9887 & 0.9837 & 0.0060 & 0.5009 & 0.9817 & 0.9990\\
5 models & 0.9883 & 0.9857 & 0.0077 & 0.4217 & 0.9800 & 0.9989\\
6 models & 0.9873 & 0.9860 & 0.0093 & 0.4007 & 0.9750 & 0.9989\\
7 models & 0.9860 & 0.9857 & 0.0087 & 0.4095 & 0.9753 & 0.9989\\
\hline
\multicolumn{7}{c}{Generalized mean ($r=-0.5$)}\\
\hline
2 models & 0.9857 & 0.9873 & 0.0147 & 0.2300 & 0.9720 & 0.9989\\
3 models & 0.9863 & 0.9823 & 0.0080 & 0.1855 & 0.9780 & 0.9985\\
4 models & 0.9848 & 0.9827 & 0.0083 & 0.1745 & 0.9780 & 0.9985\\
5 models & 0.9832 & 0.9823 & 0.0073 & 0.1848 & 0.9753 & 0.9985\\
6 models & 0.9827 & 0.9823 & 0.0100 & 0.1038 & 0.9743 & 0.9985\\
7 models & 0.9820 & 0.9837 & 0.0110 & 0.0723 & 0.9720 & 0.9985\\
\hline
\end{tabular}
\tablefoot{All metrics are evaluated on the balanced test set of 4000 simulated lenses and 4000 non-lens galaxies. The first block reports single-model performance; the second and third blocks report the arithmetic mean and generalized mean ($r=-0.5$) ensemble strategies. Accuracy (column~2) is computed at the default threshold of 0.5. TPR and FPR (columns~3 and~4) are evaluated at the threshold that maximizes Youden's $J = \mathrm{TPR} - \mathrm{FPR}$, given in column~5. Column~6 reports the true-positive rate at a fixed false-positive rate of $0.5\%$ ($\mathrm{TPR@FPR}=0.5\%$). AUC (column~7) is the area under the ROC curve, a threshold-independent summary of overall classification performance.}
\end{table*}

\begin{figure*}[t!]
\centering
\includegraphics[width=0.7\textwidth, trim=40 30 40 60, clip]{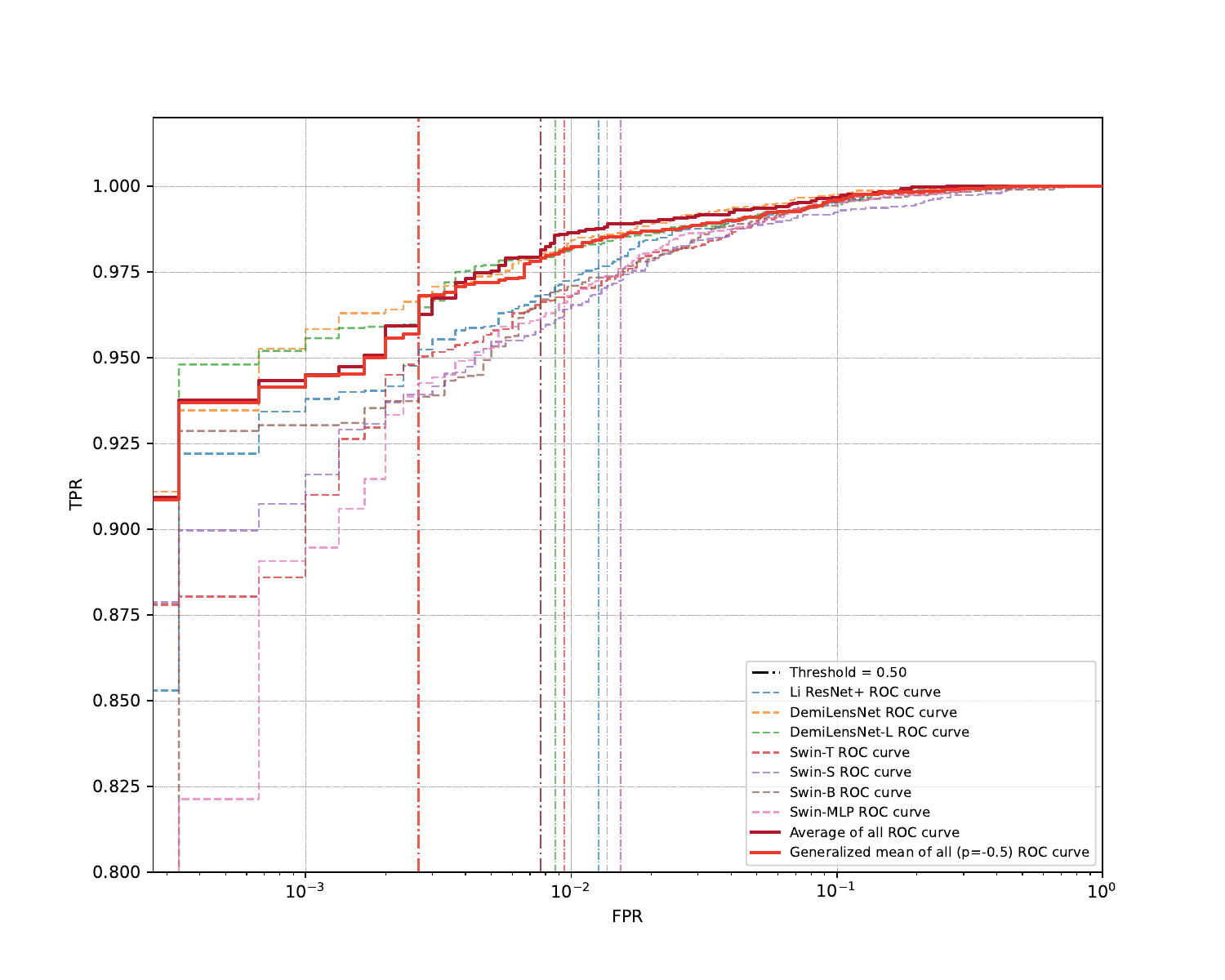}
\caption{ROC curves on the simulated test set. Dashed lines show the seven individual models; solid lines show the arithmetic mean and generalized mean ($r=-0.5$) ensemble strategies, each combining all seven models. The markers on the curves indicate the operating points corresponding to probability thresholds of 0.5.}
\label{fig:roc}
\end{figure*}

Table~\ref{tab:comparison} summarizes the performance of the individual classifiers and the two ensemble strategies on the simulated test set, which contains 4000 lenses and 4000 non-lens galaxies. As shown in the table, the models are ordered from best to weakest: DemiLensNet, Li ResNet+, DemiLensNet-L, Swin-MLP, Swin-T, Swin-S, and Swin-B. In the following, an ensemble of size $N$ consists of the first $N$ models from this sequence. The thresholds listed in Table~\ref{tab:comparison} correspond to the values that maximize Youden's \(J=\mathrm{TPR}-\mathrm{FPR}\).
This threshold provides a useful summary of the trade-off between completeness and false positives, but it should not be regarded as a fixed operational choice for a lens search.
In practice, the preferred threshold depends on the scientific goal and on the cost of false positives relative to missed lenses.
For example, increasing the threshold reduces the number of false positives but also lowers completeness, whereas decreasing the threshold recovers more lenses at the price of a larger inspection sample.
For this reason, we also report \(\mathrm{TPR@FPR}=0.5\%\), which compares the models at the same stringent false-positive rate.
This metric reduces the ambiguity associated with arbitrary threshold choices and is more directly connected to the practical problem of constructing a manageable lens-candidate list.

The best single-model performance depends on the operating regime.
DemiLensNet achieves the highest accuracy among the individual models (0.9870) and the highest AUC (0.9989), while DemiLensNet-L gives the lowest FPR at the Youden-\(J\) threshold (0.0053) and the highest \(\mathrm{TPR@FPR}=0.5\%\) (0.9770).
Thus, the two DemiLensNet variants perform best among the single models, but their advantages appear in slightly different parts of the ROC space.
Li ResNet+ is the strongest convolutional baseline in terms of accuracy, AUC, and TPR at fixed low FPR, although some Swin-based models reach lower FPR values at the Youden-\(J\) threshold.
Overall, the Swin-based models are competitive in AUC but are generally less effective than the DemiLensNet models in the stringent low-FPR regime that is most relevant for lens searches. The ensemble results show that adding more models does not necessarily improve performance on this matched simulated test set.
For the arithmetic mean, the four-model ensemble gives the best overall result, with the highest accuracy (0.9887), the lowest FPR at the Youden-\(J\) threshold (0.0060), the highest \(\mathrm{TPR@FPR}=0.5\%\) (0.9817), and the highest AUC (0.9990).
Adding further models slightly degrades these metrics.
For the generalized mean with \(r=-0.5\), the optimal subset depends on the metric: the three- and four-model ensembles give the highest \(\mathrm{TPR@FPR}=0.5\%\), whereas the five-model ensemble gives the lowest FPR at the Youden-\(J\) threshold.
The two-model generalized mean has the highest AUC among the generalized-mean combinations.
These results indicate that ensemble performance is not monotonic with ensemble size.
On the simulated test set, the best results are obtained from a small subset of strong and complementary models rather than from a simple combination of all available classifiers.
This behaviour is expected for a simulated test set drawn from the same distribution as the training data.
In this case, the different models often make highly correlated predictions, so the gain from adding more classifiers is limited.
Models with weaker low-FPR performance can dilute the ensemble score and reduce the benefit provided by the strongest classifiers.
This does not imply that larger ensembles are unhelpful in real survey applications.
For real cutouts, where the image properties, contaminants, and observational artefacts are more diverse than in the simulations, different architectures may make less correlated errors.
The ensemble can then act as a consistency filter by favouring sources that receive high lens probabilities from multiple models.

Figure~\ref{fig:roc} shows the ROC curves for the seven individual classifiers and the two seven-model ensemble scores (the arithmetic mean and the generalized mean).
All competitive models reach very high AUC values, and their ROC curves are therefore close to each other over much of the plotted range.
In the high-TPR, low-FPR regime, both ensemble curves lie among the best-performing curves, but they do not clearly outperform the strongest individual models, DemiLensNet and DemiLensNet-L.
This is consistent with Table~\ref{tab:comparison}, where the best result on the simulated test set is obtained by the four-model arithmetic mean rather than by combining all seven models.
The low-FPR range around \(10^{-3}\)--\(10^{-2}\) is particularly important for strong-lens searches, because wide-field surveys contain many more non-lenses than true lenses.
In this regime, even a small reduction in FPR can substantially reduce the number of false positives passed to visual inspection.
Since the AUC values of the strongest models are already close to unity, the practical value of the ensemble is better assessed through low-FPR quantities such as \(\mathrm{TPR@FPR}=0.5\%\), rather than through AUC alone.

\begin{figure*}
    \centering
    \includegraphics[width=\textwidth]{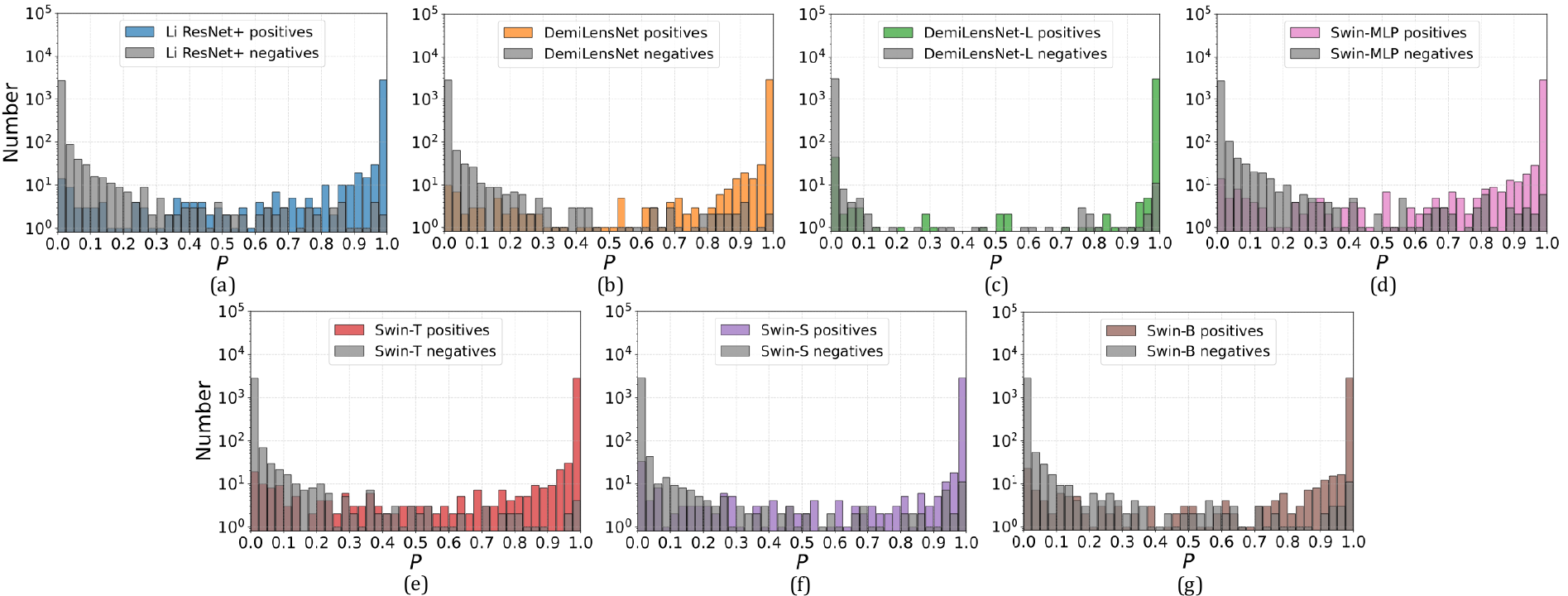}
     \caption{Predicted probability distributions for the seven individual models on the simulated test set. In each panel, the grey histogram shows the non-lens (negative) class and the coloured histogram shows the lens (positive) class. The model name is indicated in each panel.}
    \label{fig:confidence_individual}
\end{figure*}

The probability distributions in Fig.~\ref{fig:confidence_individual} and Table~\ref{table:distribution_metrics} provide a complementary diagnostic of how each classifier separates the positive and negative simulated samples.
Panels (a)--(g) show the predicted lens-probability distributions for the seven individual models, and Table~\ref{table:distribution_metrics} gives the Jensen--Shannon (JS) divergence and overlap area (OA) between the positive and negative distributions.
For these calculations, the positive and negative histograms are normalized separately into discrete probability mass functions.
The JS divergence, computed with the natural logarithm, measures the global difference between the two distributions.
The overlap area measures their shared probability mass,
\[
{\rm OA} = \sum_i \min \left[p_{+}(i),p_{-}(i)\right],
\]
where \(p_{+}(i)\) and \(p_{-}(i)\) are the normalized positive and negative distributions in probability bin \(i\), respectively.
These quantities are therefore useful diagnostics of score separation, but they should not be interpreted as direct measures of survey performance.

DemiLensNet has the largest JS divergence among the tested models, 0.6548, and a small OA of 0.0293.
DemiLensNet-L has the smallest OA, 0.0240, although its JS divergence is lower, 0.6374.
This difference is informative.
A larger JS divergence indicates stronger global separation between the lens and non-lens score distributions, whereas a smaller OA indicates less direct overlap between the two classes.
The DemiLensNet series therefore appears particularly well matched to the simulated test set.
It produces a clearer separation than Li ResNet+ and the Swin-based models, whose OAs are larger, ranging from 0.0337 for Li ResNet+ to 0.0390 for SwinTransformer-T.
This behaviour is broadly consistent with the low-FPR results in Table~\ref{tab:comparison}, especially for DemiLensNet-L.

However, a sharply separated probability distribution on simulated data is not by itself sufficient evidence of better performance on real survey images.
A highly bimodal distribution may indicate that the model has overfitted to patterns that are specific to the simulations, rather than learning features that genuinely distinguish lenses from non-lenses.
Such behaviour can be beneficial on a matched simulated test set, but it may not generalize reliably when the model is applied to real survey cutouts.
Moreover, even for a model whose separation is genuine, the practical impact depends on where the overlap between the positive and negative distributions occurs, not just on how small it is.
Overlap near high predicted probabilities is more problematic for candidate searches than overlap near the decision boundary, because it produces high-confidence false positives that are difficult to filter through visual inspection.
These two concerns---the risk of simulation-specific overfitting and the importance of overlap location---motivate the use of ensemble agreement in the subsequent real-data analysis.
Averaging the predicted probabilities across different models can reduce the influence of model-specific high-confidence errors.
The generalized mean with \(r=-0.5\) provides an even stricter form of consensus, because a single low-probability prediction can pull the ensemble score downward.
In this way, the ensemble does not rely solely on the confidence of a single model, but instead favours candidates that are consistently identified as lens-like by multiple models.
The effectiveness of this strategy is tested further on the real-data benchmark and in the full KiDS DR4 search.

\begin{table}[h!]
\caption{Distribution-separation metrics for the 7 models. }
\label{table:distribution_metrics}
\centering
\begin{tabular}{c c c}
\hline\hline
Model & JS divergence & OA \\
\hline
Li ResNet+    & 0.6433 & 0.0337 \\
DemiLensNet   & 0.6548 & 0.0293 \\
DemiLensNet-L & 0.6374 & 0.0240 \\
Swin-MLP      & 0.6374 & 0.0357 \\
Swin-T        & 0.6364 & 0.0390 \\
Swin-S        & 0.6279 & 0.0380 \\
Swin-B        & 0.6317 & 0.0367 \\
\hline
\end{tabular}
\tablefoot{The second and third columns report the Jensen--Shannon (JS) divergence and the overlap area (OA) of the predicted-probability distributions for the positives and negatives.}
\end{table}

\subsection{Performance on found existing lens candidates}
\label{sec:real_sample_test}

\begin{table}[h!]
\caption{Performance comparison for the seven models on real data.}
\label{table:real_summary}
\centering
\renewcommand{\arraystretch}{1.25}%
\begin{tabular}{c c c c}
\hline\hline
Model & Accuracy & FPR@TPR=90\% & AUC \\
\hline
\multicolumn{4}{c}{Single model} \\
\hline
DemiLensNet   & 0.9721 & 0.0200 & 0.9829 \\
Li ResNet+    & 0.9703 & 0.0160 & 0.9785 \\
DemiLensNet-L & 0.9703 & 0.0210 & 0.9823 \\
Swin-MLP      & 0.9667 & 0.0320 & 0.9814 \\
Swin-T        & 0.9640 & 0.0380 & 0.9761 \\
Swin-S        & 0.9622 & 0.0290 & 0.9772 \\
Swin-B        & 0.9631 & 0.0300 & 0.9796 \\
\hline
\multicolumn{4}{c}{Arithmetic mean} \\
\hline
2 models      & 0.9712 & 0.0080 & 0.9820 \\
3 models      & 0.9820 & 0.0100 & 0.9839 \\
4 models      & 0.9757 & 0.0110 & 0.9874 \\
5 models      & 0.9730 & 0.0100 & 0.9868 \\
6 models      & 0.9694 & 0.0150 & 0.9869 \\
7 models      & 0.9730 & 0.0110 & 0.9864 \\
\hline
\multicolumn{4}{c}{Generalized mean ($r=-0.5$)}\\
\hline
2 models & 0.9775 & 0.0090 & 0.9837\\
3 models & 0.9829 & 0.0100 & 0.9847\\
4 models & 0.9802 & 0.0100 & 0.9844\\
5 models & 0.9820 & 0.0080 & 0.9844\\
6 models & 0.9820 & 0.0080 & 0.9844\\
7 models & 0.9820 & 0.0070 & 0.9846\\
\hline
\end{tabular}
\tablefoot{All metrics are evaluated on the mixed test set of 111 LRG lens candidates and 1000 non-lens galaxies. The first block reports single-model performance; the second and third blocks report the arithmetic mean and generalized mean ($r=-0.5$) ensemble strategies. Accuracy (column~2) is computed at the default threshold of 0.5. Column~3 reports $\mathrm{FPR@TPR}=90\%$, the false-positive rate at a fixed true-positive rate of $90\%$, a commonly adopted operating point that recovers most true lenses while keeping false positives manageable. Column~4 is the area under the ROC curve (AUC).}
\end{table}

\begin{figure*}[t!]
\centering
\includegraphics[width=0.7\textwidth, trim=40 30 40 60, clip]{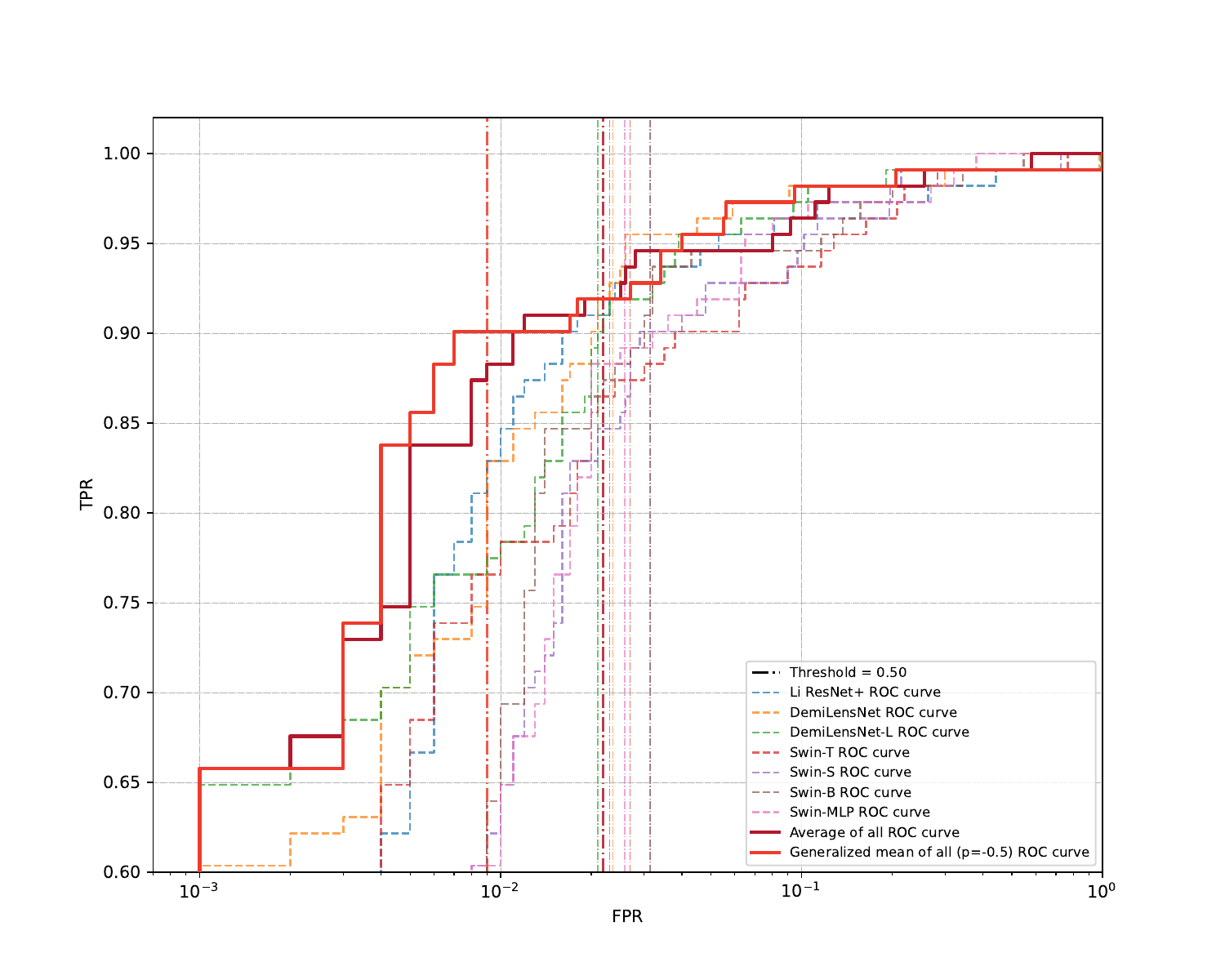}
\caption{ROC curves on real test set. Dashed lines show the seven individual models; solid lines show the arithmetic mean and generalized mean ($r=-0.5$) ensemble strategies, each combining all seven models. The markers on the curves indicate the operating points corresponding to probability thresholds of 0.5.}
\label{fig:real_roc}
\end{figure*}

\begin{figure*}[t!]
\centering
\includegraphics[width=0.9\textwidth]{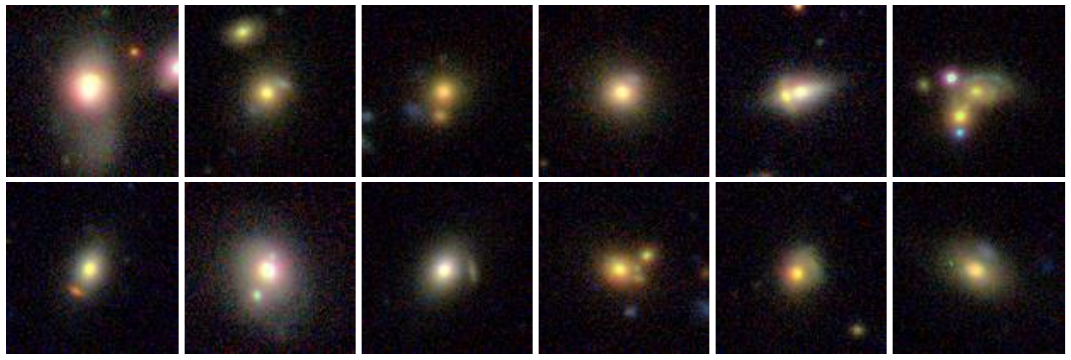}
\caption{High quality lens candidates from literature  that receive Generalized mean ensemble scores below the $\mathrm{FPR@TPR}=90\%$ threshold. Colour images are composed from the $g$-, $r$-, and $i$-band KiDS images. Most of these objects lack definitive lensing features and are morphologically ambiguous.}
\label{fig:111fp}
\end{figure*}
We next evaluate the models on real data, which provides a more realistic test of the candidate screening problem than the balanced simulated set.
The benchmark combines 111 high-quality lens candidates from the KiDS DR4 lensing sample \citep{Petrillo2019MNRAS.484.3879P, li2020} with 1000 non-lens galaxies.
We use only candidates identified in the luminous red galaxy (LRG) sample, excluding those from the bright galaxy (BG) sample (\citealt{li2020, Li_DR5lens_2021}), so that the galaxy population matches that of the simulated training data as closely as possible.
This choice isolates the effect of moving from simulated to real images, rather than confounding it with a change in the underlying galaxy population.

The ranking of the individual models shifts only slightly compared with the balanced simulated set (Table~\ref{table:real_summary}).
DemiLensNet again attains the highest accuracy (0.9721) and AUC (0.9829), while Li ResNet+ gives the lowest false-positive rate at 90\% completeness (FPR@TPR\(=90\%\) = 0.0160).
This drop in overall performance relative to the simulated data---for example, the best single-model FPR rises from sub-percent values to \(1.6\%\)---is expected, because the models were trained on simulations and now face the full morphological complexity of real galaxies.
The stability of the ranking nevertheless indicates that the models which perform best on simulations remain competitive on real data. The two ensemble strategies respond very differently as the ensemble grows, and this contrast is the central result of this section.
For the arithmetic mean, the false-positive rate at fixed completeness does not improve steadily with ensemble size: the two-model average already reaches FPR = 0.0080, but adding weaker models tends to drive the FPR back up, reaching 0.0150 for six models.
The reason is that the arithmetic mean gives every model equal weight, so a confidently assigned but incorrect high score from a weaker model is incorporated directly into the ensemble score and can push a non-lens above the threshold.
The generalized mean with \(r=-0.5\) shows the opposite behavior: its false-positive rate trends downward as the ensemble grows, in contrast to the erratic fluctuations of the arithmetic mean, ultimately reaching its lowest value with all seven models (FPR = 0.0070, i.e. 7 false positives out of 1000 non-lenses).
This follows directly from the property of the generalized mean with a negative exponent: a single low probability pulls the combined score downward.
A non-lens is therefore classified as positive only if \emph{all} models assign it a high score; a high score from one overconfident model is no longer sufficient.
The generalized mean thus acts as a consensus filter, and adding more models strengthens rather than dilutes this filter, because each additional model provides another opportunity to veto a spurious detection.

This behavior is the opposite of the trend seen on the simulated benchmark, where increasing the ensemble beyond a few models produced no further reduction in the false-positive rate because the individual predictions were highly correlated.
On real data, the errors made by individual models are more diverse and less strongly correlated, and the consensus requirement of the generalized mean is what allows this diversity to translate into a lower false-positive rate.
Figure~\ref{fig:real_roc} illustrates this contrast.
On the simulated benchmark, the best-performing curve in the low-FPR regime was produced by a single model, DemiLensNet.
On the real-data benchmark, the two uppermost curves in this regime are instead the ensemble scores, with the generalized mean lying above the arithmetic mean.
Both ensembles sit clearly above all individual models for \(\mathrm{FPR} < 10^{-2}\), the range most relevant for candidate screening.
The change in which curve is best---from a single model on simulations to the ensemble on real data---is the clearest evidence that the practical value of the ensemble approach becomes apparent only when the models are exposed to the morphological diversity of real observations.

To examine which candidates the generalized mean rejects, Fig.~\ref{fig:111fp} shows the literature candidates whose ensemble scores fall below the FPR@TPR\(=90\%\) threshold.
These objects are treated as positives in the evaluation, so excluding them is counted as a loss of completeness: the ensemble is penalised in the metric for suppressing them.
On visual inspection, however, most of these objects do not display convincing lensing features.
Several show possible arc-like structures, but their morphologies are equally consistent with galaxy pairs or with substructure in the central galaxy. 
This distinction has an important implication.
The low ensemble scores do not simply correspond to genuine lenses that the ensemble has missed.
Rather, the consensus requirement of the generalized mean assigns lower scores to morphologically ambiguous systems, several of which are likely contaminants in the original catalogues.
Part of the apparent penalty in the completeness metric therefore reflects the ensemble correctly filtering out candidates that are probably not real lenses, and the actual false-positive suppression is likely somewhat better than the nominal FPR values suggest.
This behaviour is precisely the kind of behaviour needed for wide-field searches, where the dominant cost is the visual inspection of large numbers of ambiguous, marginally lens-like objects.

\section{Applying to KiDS DR4}
\label{sec:apply_to_dr4}

\subsection{Predictive Data}
We constructed two distinct predictive samples for strong gravitational lens identification using our CNN models: Bright Galaxies (BGs) and LRGs. These samples were derived from the KiDS DR4 catalog, following selection thresholds established in previous works (\citealt{li2020, Li_DR5lens_2021}). The KiDS-DR4 catalog, comprising approximately 120 million detected sources, served as our initial dataset.

Our first predictive sample, the BG sample, was derived from the complete KiDS-DR4 galaxy catalog. This sample was designed for broad coverage, aiming to include a large number of potentially lensing galaxies. To achieve this, we selected objects that met two primary criteria: first, they were classified as galaxy-like objects, indicated by the \texttt{SG2DPHOT} flag being set to 0, a flag generated by the \texttt{2DPHOT} software during KiDS catalog extraction. Second, their $r$-band Kron-like magnitude (\texttt{mag\_auto} from \texttt{Sextractor}) was required to be $r_{\text{auto}}\leq21$. These criteria yielded a BG sample of 3,909,523 galaxies. This number represents a slight increase compared to the number (3,808,963) reported in earlier studies utilizing KiDS DR4. This minor discrepancy is attributed to the fact that the KiDS team later identified processing issues within more than 30 tiles of the original DR4. Consequently, a reprocessed DR4.1 catalog was released, and we incorporated these updated catalogs for the affected tiles, leading to the slightly larger final BG sample size. Derived directly from the BG sample, our second predictive sample, the LRG sample, focuses on a more refined subset of galaxies. These LRGs are inherently more likely to act as strong lenses due to their characteristically higher masses. The LRG sample is thus a subset of the BGs, further refined by specific color-magnitude cuts. We adopted the established approach of \cite{Petrillo2019MNRAS.484.3879P}, adapting the low-redshift ($z < 0.4$) color-magnitude selection originally outlined by Eisenstein et al. (2001). This involved applying additional criteria based on the \texttt{COLOUR\_GAAP} color indices ($g-r$ and $r-i$) available in the KiDS-DR4 catalog. This refined selection process yielded an LRG sample of precisely 146,960 galaxies.

\subsection{New candidates}
\label{sec:new_candidates}

\begin{figure}[h!]
\centering
\includegraphics[width=\columnwidth]{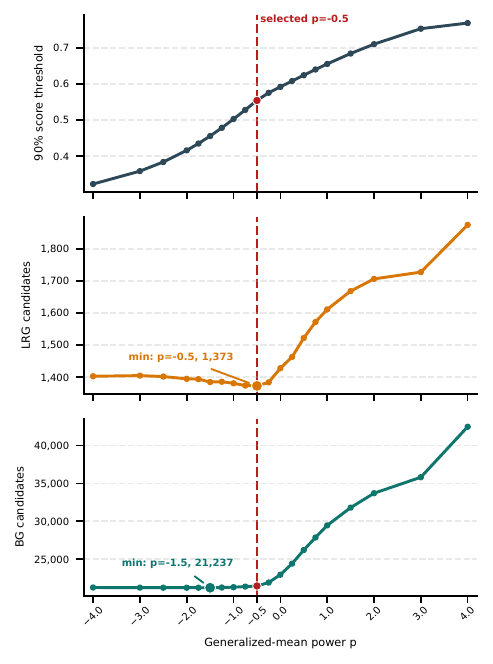}
\caption{Sensitivity of the seven-model generalized-mean ensemble to the power $r$ at 90\% completeness. From top to bottom, the panels show the score threshold, the number of selected LRG candidates, and the number of selected BG candidates.}

\label{fig:p_sensitivity}
\end{figure}
The exponent $r$ of the generalized mean is chosen to minimise the number of candidates returned at 90\% completeness on the real LRG sample. Although $r$ could in principle be set using the simulated data, the models are ultimately applied to real survey images, and selecting $r$ on real data ensures that the choice is tuned to the target domain. Figure~\ref{fig:p_sensitivity} shows how the score threshold and the resulting candidate counts vary with $r$ for the seven-model ensemble.
The upper panel gives the score threshold at 90\% completeness, determined from the 111 LRG benchmark lenses.
The middle and lower panels show the corresponding numbers of LRG and BG candidates.
In general, both curves fall as $r$ becomes more negative, reflecting the increasingly strict consensus requirement, reach a minimum, and then remain nearly flat over a broad range of negative $r$ before rising sharply for $r > 0$. We adopt $r=-0.5$, where the LRG candidate count reaches its minimum of 1373.
This choice is based on the LRG sample because it is our primary search population and directly sets the completeness calibration.
The BG trend supports this choice: the BG minimum occurs at $r=-1.5$ (21,237), and at $r=-0.5$ the count is only slightly higher (21,472, see Table~\ref{tab:model_performance}), so the LRG-optimal $r$ is near-optimal for the BG sample as well.
The steep rise in both curves at positive $r$ confirms that simply taking a standard average ($r=1$) or emphasising the highest individual scores would inflate the candidate lists, undoing the consensus filter that negative $r$ provides.

\begin{table}[h!]
\caption{Number of candidates recovered at $\mathrm{TPR}=90\%$ on the LRG benchmark for the full LRG and BG samples.}
\label{tab:model_performance}
\centering
\renewcommand{\arraystretch}{1.25}%
\begin{tabular}{c c c c}
\hline\hline
Group & 90\% threshold & LRG & BG \\
\hline
\multicolumn{4}{c}{Single model} \\
\hline
DemiLensNet   & 0.6016 & 2715 & 69843  \\
Li ResNet+    & 0.6103 & 3067 & 93265  \\
DemiLensNet-L & 0.4757 & 2939 & 95518  \\
Swin-MLP      & 0.3959 & 4402 & 133548 \\
Swin-T        & 0.3538 & 4562 & 117706 \\
Swin-S        & 0.5331 & 4097 & 97321  \\
Swin-B        & 0.5363 & 3966 & 114244 \\
\hline
\multicolumn{4}{c}{Arithmetic mean} \\
\hline
2 models & 0.6703 & 1892 & 41972 \\
3 models & 0.6105 & 1912 & 44341 \\
4 models & 0.5979 & 1893 & 42047 \\
5 models & 0.6231 & 1685 & 33807 \\
6 models & 0.6264 & 1718 & 32842 \\
7 models & 0.6555 & 1611 & 29460 \\
\hline
\multicolumn{4}{c}{Generalized mean ($r=-0.5$)}\\
\hline
2 models & 0.6482 & 1860 & 40966\\
3 models & 0.5669 & 1726 & 37788\\
4 models & 0.4827 & 1740 & 35525\\
5 models & 0.4721 & 1649 & 29467\\
6 models & 0.5161 & 1496 & 24935\\
7 models & 0.5544 & 1373 & 21472\\
\hline
\end{tabular}
\tablefoot{The threshold (column~2) is the probability above which 90\% of the 111 LRG benchmark lenses are recovered; it is determined independently for each model or ensemble and then applied unchanged to the full LRG and BG samples. Columns~3 and~4 report the total number of candidates returned above that threshold. These totals include both genuine lenses and false positives.}
\end{table}

We applied all seven models to both the LRG and BG samples to predict the lensing score for each galaxy. We first examine the predictions of each model on the LRG sample. Table~\ref{tab:model_performance} shows the threshold at a target completeness of 90\%, and the total number of predicted candidates above this threshold. Among the individual models, DemilensNet performs best, yielding only 2,715 candidates above the threshold. ResNet, which was widely used in our previous work, predicts a slightly larger number, indicating a correspondingly higher false positive rate (FPR). The Swin series models predict approximately 4,000 candidates under the same completeness requirement.
We then evaluate the performance of model ensembles. In the same table, we present the average scores and the generalized mean scores of different model combinations. Simply averaging the scores helps reduce the FPR; the combination of all seven models provides the best prediction, with 1,611 candidates at 90\% completeness. This represents a reduction of approximately 41\% compared to the best single model. Furthermore, when the generalized mean score is used, the FPR is further reduced to 1,347 candidates—a reduction of about 50\%—indicating that the generalized mean score outperforms simple averaging.
For the BG sample, the best single model is still DemilensNet, which predicts 69,843 candidates. Averaging the seven models reduces this number to 29,460. Using the generalized mean score further reduces it to 21,472, which is only one-third of the number predicted by the best single model and only one-quarter of that predicted by the commonly used ResNet.
Comparing the performance under the generalized mean scores with our previous work, the false-positive fraction decreased significantly. In \cite{li2021}, we used the same training data as in the current study. Applying the network to the LRG sample at the same 90\% completeness threshold, the three-band version of the network yielded a false-positive fraction of 1.8\%. Here, the false-positive fraction is 0.93\%, corresponding to a reduction of about 50\%. For the BG sample, using the same threshold, the previous network predicted more than 90{,}000 candidates, whereas the ensemble predicts only $\sim$21,472 candidates, i.e., roughly one-fourth of the previous number. This reduction makes it feasible to perform visual inspection for all candidates.

\begin{figure*}[t!]
\centering
\includegraphics[width=0.9\textwidth]{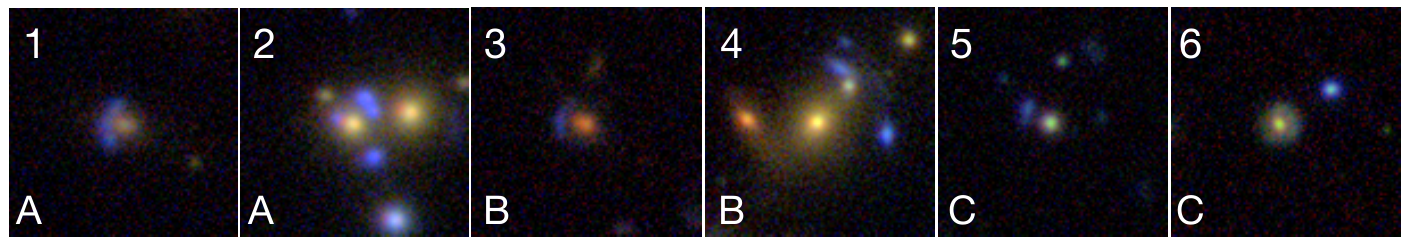}
\caption{Representative examples of the three classification classes. Panels~1--2: Class~A; panels~3--4: Class~B; panels~5--6: Class~C. Colour images are composed from the $g$-, $r$-, and $i$-band KiDS images.}
\label{fig:sample}
\end{figure*}

\begin{figure*}[t!]
\centering
\includegraphics[width=\textwidth]{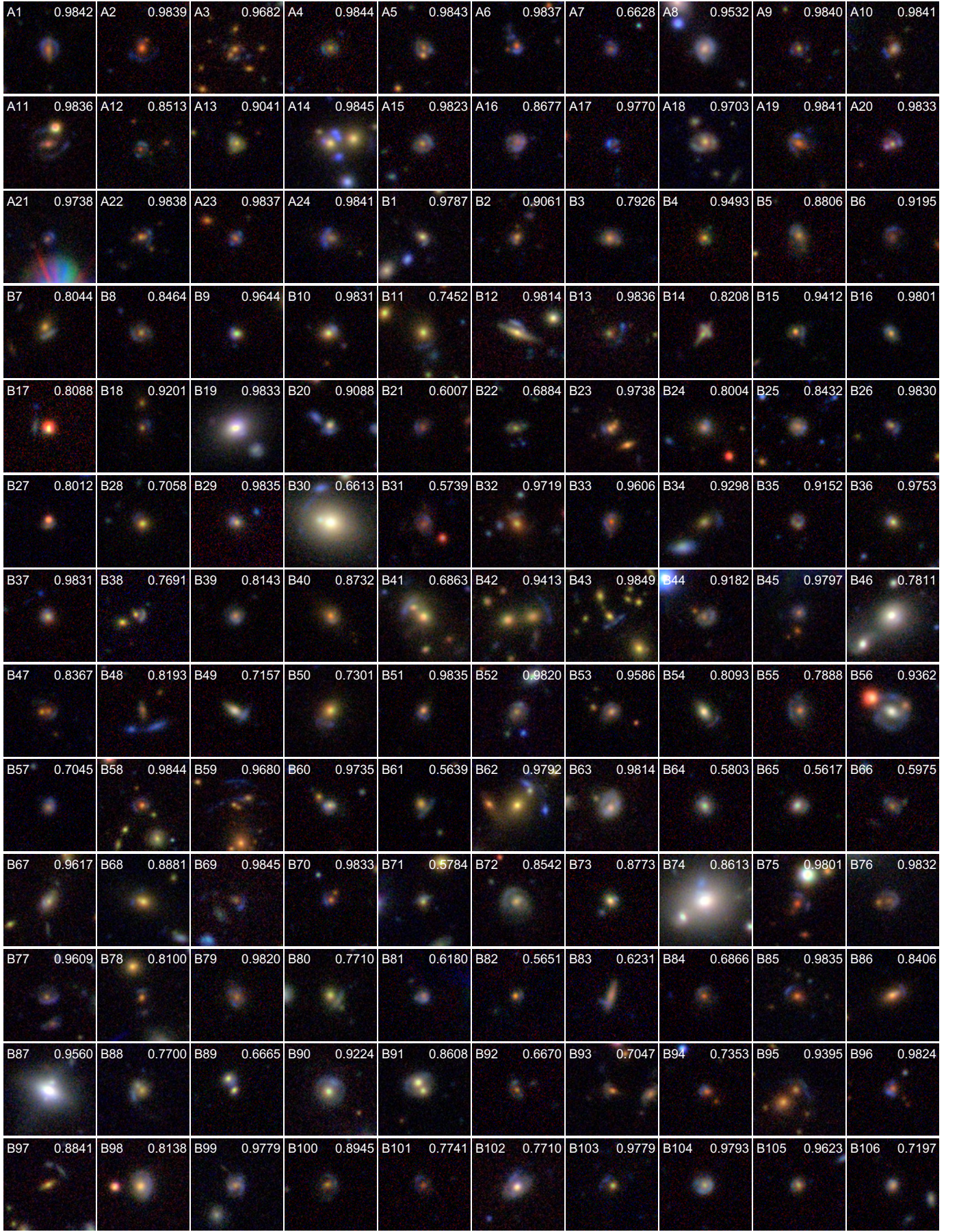}
\caption{Color stamps of Class A and Class B candidates, created by combining the $g$, $r$, and $i$-band KiDS images. For each candidate, the generalized-mean score from the multi-models is shown at the upper right. The first 24 stamps show Class A candidates, and the remaining ones show Class B candidates.}
\label{fig:candidates}
\end{figure*}

\begin{figure*}[t]
\ContinuedFloat
\centering
\includegraphics[width=\textwidth]{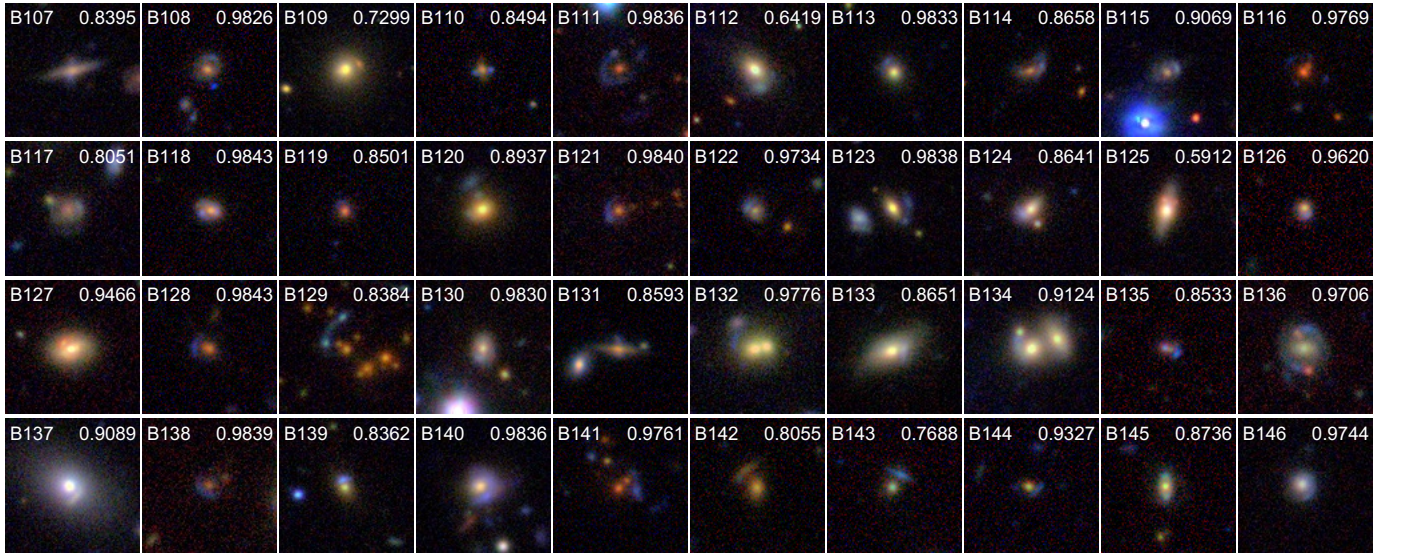}
\caption{Continued.}
\end{figure*}

We first remove 365 previously identified lens candidates from the LRG and BG samples, leaving 21,107 candidates. The sample size remains large while expert manpower is limited. We therefore assign three inspectors (ZL, HL, and XH) to flag objects that clearly do not show lens features. Such contaminants often include spiral galaxies, mergers, pairs, and other morphologies that are unambiguously not gravitational lenses. After this step, 1876 candidates remain. Then, we assign a single expert (RL) to classify the remaining candidates into three classes:
\begin{description}
    \item[Class A:] 
   As shown in the first two cases of Fig.~\ref{fig:sample}, these candidates provide the strongest evidence for gravitational lensing. Morphologically, they exhibit clear multiple images that are aligned tangentially with respect to the center of the deflector. In these systems, the lensed sources often appear as blue, star-forming galaxies, which produces a sharp contrast with the reddish central massive elliptical foreground galaxy. These candidates are usually straightforward to model, allowing the foreground mass distribution to be constrained from the multiple lensed images.

    \item[Class B:] 
    As shown in the middle two cases of Fig.~\ref{fig:sample}, these systems are characterized by giant arcs or multiple lensed images, but they may lack one or more counter images needed to form a complete lens configuration. Nevertheless, they remain plausible genuine lenses, as incompleteness can arise from limited image quality or insufficient detection of some counter images. We classify these objects as high-probability lens candidates, provided that they are confirmed through follow-up spectroscopy.

    \item[Class C:]
   The remaining objects are assigned to Class C. Although these candidates may represent lensed images of background sources, they can also arise from other physical phenomena (see Fig.~\ref{fig:sample}), such as tidal debris, faint satellite galaxies, or ring galaxies—or the lensed arcs may have too low S/N to be clearly resolved. We retain these objects in the catalog, but their purity is expected to be substantially lower than that of Classes A and B. These samples can also be prioritized for detailed inspection in future fourth-generation sky surveys. 
\end{description}
Finally, we obtain 24 Class A, 146 Class B, and 1706 Class C new lens candidates. Both Class A and Class B candidates show strong evidence of lensing and are therefore well suited for spectroscopic follow-up to determine source redshifts. We collectively refer to these two classes as high-quality lens candidates, and shown their color images in Fig. \ref{fig:candidates}. Within this high-quality sample, Class A corresponds to the high-confidence subset—the most secure candidates—while Class B corresponds to the likely lens subset. For most of the Class B candidates, parts of the contour images are difficult to discern; nevertheless, even when followed up with telescope observations, they still show a high probability of being genuine lenses. In Fig.~\ref{fig:sample}, we present the newly identified Class A and Class B candidates.

\subsection{Assembled candidate Catalog in KiDS from Multiple works}
Many studies related to strong-lensing searches have been carried out using KiDS data, and different groups have identified candidates using different pipelines (\cite{2017MNRAS.472.1129P, Petrillo2019MNRAS.484.3879P, li2020, Li_DR5lens_2021, 2023MNRAS.523.4188N, 2024MNRAS.533.1426N}). However, because these works were performed independently, there is no single uniform definition of candidate quality. In this work, we compile all previously reported candidates and the ones found in this work, and classify them into three classes (Class A, Class B, and Class C) according to the criteria described above. This yields 91 Class A, 314 Class B, and 1892 Class C candidates, which together represent all strong-lensing candidates found in the KiDS survey to date. \cite{2017MNRAS.472.1129P} predicted that approximately 2,400 lenses could be detected in the $\sim 1,300$ $deg^2$ KiDS survey. This figure is close to the total number of candidates we have compiled. However, this does not necessarily mean that all lenses in KiDS have been detected, as a large number of false positives in Class C would be removed if high-quality imaging and spectroscopic follow-up observations were conducted. Therefore, many candidates in KiDS likely remain undiscovered. 

We construct a final sample based on this combined catalog. The resulting catalog, including RA, Dec, class, and the corresponding images, is available at \url{https://cosviewer.com/datasets/strong-lens-kids}. Candidates newly identified in this work are marked with an asterisk ($*$) after the class.

\section{Discussion\label{sec:discussion}} 

This work no longer aims at improving the accuracy of an already existing single network, but at suppressing the false positive rate when the method is applied to realistic survey data. This distinction matters because strong-lens searches are fundamentally constrained by the rarity of lenses in the real world: the decisive question is not only how many true lenses are recovered, but how many contaminants survive into the human-vetting stage \citep{li2020, li2021, denselens2023, 2025MNRAS.538.1081R}. In this context, the false positive rate at a fixed true positive rate of 90\% drops from 0.018 for the best single real-data model (e.g., \cite{Li_DR5lens_2021}) to 0.009 for our ensemble of seven models—a reduction that is operationally significant. Under a simple linear scaling, this corresponds to roughly 9,000 fewer contaminants per million galaxies at fixed completeness ($\sim 90\%$ in this work). For candidate-ranking pipelines, where visual inspection time and follow-up resources are the limiting factors, this is the more relevant metric.

The ensemble operation of multiple models further points to a specific mechanism rather than a purely generic averaging effect. Galaxy-scale lenses present a comparatively coherent set of cues—curved tangential light, partial rings, or multiply imaged arc segments around a foreground deflector—that can be recognized across different architectures. False positives, by contrast, are morphologically heterogeneous: spiral arms, mergers, ring galaxies, residual subtraction artifacts, and chance superpositions need not trigger the same response in different models. In this sense, averaging acts as a consistency filter. Contaminants are suppressed not simply because variance is reduced in an abstract statistical sense, but because many pseudo-lenses fail to elicit stable cross-model agreement. This interpretation is consistent with previous lens-finding studies showing that the reduction of false positives depends both on representative training data and on ensemble strategies \citep{denselens2023, 2025MNRAS.538.1081R}.

Ensembles can also suppress the impact of discrepancies between simulations and real data. Most of the lensing search networks are trained on simulated data. In fact, simulations cannot model all observational effects and irregular structures of galaxies, creating inherent differences between simulated and real data. When moving from simulation to real survey images, the data may encode domain shift, atypical morphology, or other forms of sample difficulty. This interpretation is consistent with the broader literature on strong-lens classification, which has repeatedly shown that the gap between simulated positives and the contaminant population present in survey data remains a major limitation of current methods \citep{rojas2022, denselens2023, 2025MNRAS.538.1081R}. This improvement can be understood as a consequence of the diverse inductive biases across architectures. Since individual models are trained exclusively on simulated data, they inevitably overfit to simulation-specific features (e.g., simplified background subtraction, regular galaxy morphologies) in different ways. When applied to real survey data, these overfitted features lead to distinct patterns of false positives: for example, a CNN-based model may be more sensitive to high-curvature local features such as spiral arms, whereas a transformer-based model with global attention may be more robust to such local features but more susceptible to extended low-surface-brightness artifacts. By making together the predictions of multiple architectures, the ensemble selectively retains candidates that elicit consistent responses across all models, effectively requiring cross-architectural agreement. This mechanism naturally suppresses simulation-driven false positives that stem from architecture-specific overfitting rather than from genuine lens features.

Looking ahead to future wide-field surveys, the challenge of false positives becomes a dominant concern. Surveys such as ESA's \textit{Euclid} mission and the Vera C. Rubin Observatory's Legacy Survey of Space and Time (LSST) will image billions of galaxies and are expected to discover on the order of $10^5$ new strong lenses \cite{Collett_2015}. At this scale, even a modest false positive rate leads to an overwhelming number of contaminants. For example, applying a single model with a false positive rate of 1.8\% to one billion galaxies would yield approximately 18 million false candidates. Human inspectors cannot feasibly vet such a large sample, making manual inspection the primary bottleneck in the lens discovery pipeline rather than the detection itself \cite{denselens2023, 2025MNRAS.538.1081R}. Reducing the false positive rate is therefore not merely a matter of statistical improvement, but a prerequisite for enabling survey-scale lens searches.
Ensemble strategies offer a practical path toward this goal. In this work, our ensemble of seven models reduces the false positive rate by approximately 50\% compared to the best single model (from 1.8\% to 0.9\%; see Sect.~\ref{sec:new_candidates}). Extrapolating to the billion-galaxy scale, this corresponds to a reduction from roughly 18 million false positives to 9 million, sparing approximately 9 million candidates from human inspection. While running multiple models instead of a single one incurs additional computational cost, the savings in expert visual inspection time far outweigh this overhead. The extra compute is a one-time, automated expense, whereas manual vetting is slower, more expensive, and inherently limited; citizen-science projects such as Space Warps can further reduce the expert workload through candidate pre-screening \citep{Marshall2016}. The trade-off is therefore strongly favorable: the ensemble's computational cost is modest compared to the human effort it saves.

Nevertheless, even with this improvement, a false positive rate of 0.9\% applied to billions of galaxies would still yield millions of false candidates requiring visual inspection. The candidate list, although substantially cleaner than that of a single model, remains too large for exhaustive manual vetting. Further reductions in false positives will be necessary before automated lens searches can operate at survey scale with minimal human intervention. Future work may explore more advanced ensemble strategies---such as weighted averaging based on per-model confidence, hierarchical classification cascades, or integration with morphological priors---to push the false positive rate lower while maintaining high completeness. The results presented here demonstrate that cross-architectural ensemble is a promising direction, but continued development of ensemble methods, alongside improvements in training data and simulation realism, will be essential for realizing the full scientific potential of upcoming surveys. Beyond architectural improvements, the training data itself offers substantial room for progress. On the positive side, more realistic lens simulations—incorporating complex source morphologies, dust extinction, and realistic point-spread-function convolution—can produce training samples that better capture the diversity of real gravitational lenses. Recent Euclid results also demonstrate the value of augmenting simulation-based training with real lenses and non-lenses to better specialize the classifier to the target survey \citep{2026A&A...711A..32E}. On the negative side, collecting more representative non-lens examples from real survey data (e.g., ring galaxies, tidal features, and merger remnants) and incorporating them into training can significantly improve a model's ability to reject common contaminants. Furthermore, the coming generation of space-based surveys—Euclid, the Nancy Grace Roman Space Telescope, and CSST—will provide intrinsically sharper, higher-resolution imaging across wide fields. With improved image quality, the morphological distinction between true lensed arcs and impostors becomes cleaner, naturally reducing the false positive rate regardless of the underlying model architecture. These developments, combined with more effective ensemble strategies, point toward a future where automated lens searches can deliver samples that are both complete and clean enough for detailed statistical studies.

\section{Conclusion\label{sec:conclusion}} 
We have presented a multi-model ensemble strategy designed to reduce false positives in strong-lens searches in wide-field surveys. By combining seven classifiers with different architectures — including convolutional networks, Transformers, and hybrid models — through a generalized mean consensus, we significantly reduce the false positive rate at fixed completeness. On real KiDS DR4 data, the false positive rate at 90\% completeness drops from 0.016–0.020 for the best individual models to 0.007 for the ensemble. When applied to KiDS data, this reduction makes a practical difference: the number of predicted candidates for the LRG sample is cut by roughly 50\%, and for the BG sample by roughly 75\%. After visual inspection, we obtain 170 new high-quality candidates, including 24 Class A and 146 Class B. We have also collected lens candidates from previous studies and assembled a unified catalog of all KiDS DR4 strong-lens candidates found to date, containing 91 Class A, 314 Class B, and 1,892 Class C objects.

This strategy is especially valuable for Stage-III wide-field surveys such as KiDS, DES, and HSC, which typically contain millions of galaxies. The multi-model ensemble yields thousands of candidates---a volume that can still be visually inspected by human reviewers. However, for Stage-IV surveys such as Euclid, CSST, and Roman, the absolute number of false positives at the billion-galaxy scale remains too large for purely human screening. Recent Euclid Q1 lens searches have therefore adopted hybrid workflows combining machine learning, citizen science, expert vetting, and ensemble classification \citep{2026A&A...711A..26E,2026A&A...711A..28E,2026A&A...711A..30E}. Further advances are still needed, including more effective ensemble strategies, more realistic lens simulations, and more representative non-lens samples. In the longer term, the higher image quality expected from space-based surveys will also help, as sharper images naturally make the distinction between true arcs and impostors clearer. Reducing the false-positive rate to a level where extensive human vetting is no longer needed remains an open challenge, and one that the community will need to address as we move into the era of billion-galaxy surveys.

\bibliographystyle{aa}
\bibliography{aa}

\end{document}